\documentclass[fleqn,10pt]{wlscirep}
\usepackage[utf8]{inputenc}
\usepackage[T1]{fontenc}
\usepackage{orcidlink}

\title{Scaling laws for rockfall impact fragmentation emerging from diverse lithologies}

\author[1,*]{Álvaro Vergara \orcidlink{0009-0003-4257-1442}}
\author[2]{Sergio Palma \orcidlink{0000-0003-2963-7390}}
\author[1]{Raúl Fuentes \orcidlink{0000-0001-8617-7381}}

\affil[1]{Chair of Geotechnical Engineering and Institute of Geomechanics and Underground Technology, RWTH Aachen, Aachen 52074, Germany}
\affil[2]{Department of Mining Engineering, Metallurgy and Materials, Universidad Técnica Federico Santa María, Santiago 8940897, Chile}
\affil[*]{alvaro.vergara@rwth-aachen.de}

\begin{abstract}
Impact-induced fragmentation is a fundamental dissipative process in geosciences, yet its stochastic nature makes predicting debris evolution a persistent challenge. Here, we introduce a discrete element framework to resolve fragmentation mechanics across a diverse lithological spectrum—from high-strength siliciclastic units to massive carbonates—validated against high-resolution field data from documented rockfall events. Our results reveal that, despite the inherent randomness of impact dynamics, fragment size distributions consistently follow a universal Weibull scaling law, independent of lithology or initial kinetic energy. By applying a relative breakage index, we demonstrate a remarkable collapse of fragmentation data onto a single statistical signature, bridging the gap between grain-scale fracture and macroscopic debris evolution. We find that this Weibullian signature acts as a proxy for lithological sensitivity, reflecting distinct efficiencies in converting kinetic energy into new fracture surfaces. This framework explicitly resolves the energy partitioning between surviving blocks and comminuted debris, providing a robust predictive link between impact mechanics and structural resilience. From an engineering perspective, our findings enable a shift from idealised single-block impact assumptions toward a realistic assessment of distributed energy in fragmented particle clouds, offering a physical basis for optimising protective galleries and hazard mitigation strategies in complex mountainous terrains.

\end{abstract}
\begin{document}

\flushbottom
\maketitle
\thispagestyle{empty}

\noindent Impact-induced fragmentation represents a fundamental mechanism of material disaggregation across the physical sciences, playing a pivotal role in fields as diverse as industrial comminution, planetary impacts, and the evolution of geological landscapes. In the context of brittle geo-materials, this process is characterised by an abrupt and complex transformation of kinetic energy into new fracture surfaces, a transition governed by the stochastic distribution of internal micro-defects. Despite the inherent randomness and the vast range of scales involved, a unifying feature of these high-strain-rate events is the emergence of reproducible statistical regularities in the resulting fragment size distributions \cite{Wittel2008}. Characterising the mechanics and scaling laws that underpin this emergence remains a central challenge in geophysics; it is essential not only for deciphering the energy budget of mass movements but also for enhancing the predictive capacity of hazard models. While deterministic approaches often struggle to capture the multi-scale nature of rockfall fragmentation, the integration of discrete mechanical simulations and statistical thermodynamics offers a promising path to reconcile local fracture processes with global particle size evolution \cite{Einav2007b}.

The transition from theoretical models of brittle failure to the stochastic reality of natural slopes requires a robust understanding of how geological inheritance and impact dynamics interact. This is particularly relevant in gravity-driven mass movements, where fragmentation acts as a primary dissipative mechanism, fundamentally altering the energy balance of the system. By bridging the gap between small-scale fracture mechanics and large-scale geomorphic evolution, it becomes possible to treat these events not as isolated accidents, but as predictable physical processes governed by intrinsic scaling laws.

\begin{figure}[htbp]
\centering
\includegraphics[width=0.4\linewidth]{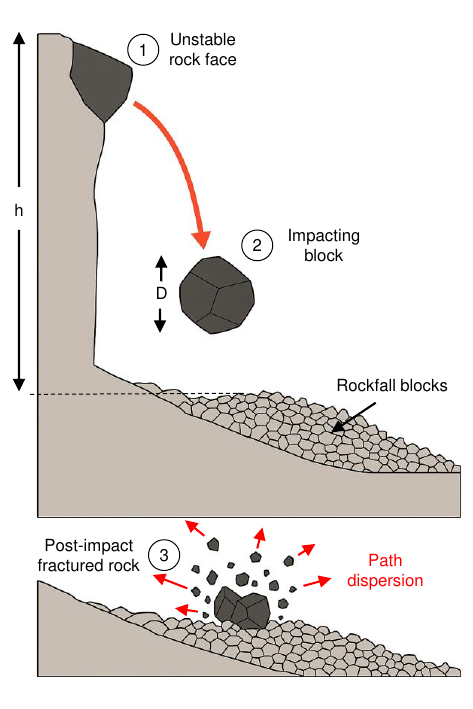}
\caption{General diagram of a single rockfall on steep slopes. In this schematic, $h$ denotes the impact height of the in-situ rock block, and $D$ its initial characteristic size. The rockfall sequence begins with the recognition of the unstable block on the slope \textcircled{\scriptsize 1}. Once it detaches from the face of the rock mass it falls freely and impacts the sediment layer \textcircled{\scriptsize 2}. Depending on the degree of energy of the collision, this impact can generate the rebound of the impacting block, or its fragmentation into several fragments \textcircled{\scriptsize 3}, where each one can follow a dispersion of different trajectories downstream, depending on the mechanical energy of the system, and experience new fragmentation events.}
\label{fig:scheme}
\end{figure}

Within this framework, rockfalls represent a primary agent of brittle fragmentation in mountainous terrain (Fig. \ref{fig:scheme}). In steep bedrock environments, these events involve the rapid detachment and downslope transit of discrete rock masses, typically mobilising volumes up to 10$^5$ m$^3$ \cite{Hungr2014}. Beyond this threshold, the process can transition into a rock avalanche, a far more destructive phenomenon characterised by the disaggregation of volumes exceeding 10$^7$ m$^3$ \cite{DeBlasio2014}. For the rock mechanics practitioner, this distinction represents a fundamental shift in dynamics: from the motion of isolated blocks to a complex, fragmenting flow where energy dissipation is dominated by internal breakage \cite{DeBlasio2015}. Understanding the transition between these regimes is essential for developing reliable run-out models and quantifying the geomorphic impact of rock slope failures.

Rockfalls drive hillslope denudation and control the pace of cliff retreat, carving persistent topographic signals into mountain landscapes \cite{Borella2019,Mohadjer2020}. High-resolution reconstructions of past events reveal that individual rockfalls can suddenly reshape alpine terrain and reconfigure sediment fluxes \cite{Caviezel2021}. Climate change is amplifying these processes: warming-induced fracturing and permafrost degradation are increasing the frequency of rockfall events, with seasonal analyses pointing to heightened activity under warming scenarios \cite{Collins2016,Stoffel2024}.

A key control on these dynamics is the impact-driven fragmentation of falling blocks. By breaking apart over 60\% of the original mass, this process enhances the mobility of finer debris and extends the run-out across multi-kilometre paths \cite{Bowman2012}. These effects leave measurable geomorphic imprints: block-size distributions and surface roughness not only encode the magnitude of past events but also serve as predictive indicators of slope instability  \cite{Barlow2012,Krautblatter2012}.

Despite this importance, the physics of rock fragmentation during impact remain incompletely understood. Field studies and experiments have shed light on some key variables: for example, the angle of impact has been shown to strongly influence the degree of fragmentation, with steeper trajectories leading to more intense breakage \cite{Giacomini2009}. Empirical observations from natural rockfall events indicate that approximately 43\% of impacting blocks fragment upon collision, generating fragment size distributions that commonly follow power-law scaling, reflecting the scale-invariant nature of brittle fragmentation \cite{Dussauge2002,Dussauge2003,Ruizcarulla2015,Gili2016}.

Discrete element models have revealed how concentrated contact forces trigger fragmentation cascades and shape run-out patterns \cite{Crosta2002,Thoeni2014,Zhao2017,Zhao2018}. Further studies have shown that fragmentation efficiency is sensitive to projectile geometry, impact kinematics, and substrate properties-highlighting the complexity of energy dissipation at impact \cite{Breugnot2016,Shen2019,Shen2020}. Collectively, these efforts have underscored the need for physically grounded models to inform hazard mitigation strategies in mountainous terrain \cite{Labiouse1996,Dorren2003,Volkwein2011,Ferrari2016}.

This study introduces a robust numerical framework based on the Discrete Element Method (DEM), designed to isolate and quantify the mechanical controls on impact-driven rock fragmentation. Central to our approach is the grounding of numerical simulations in three high-resolution case studies from Catalonia, Spain, which serve as a diverse natural laboratory. These events, spanning from the high-relief environments of the Eastern Pyrenees to the Catalan Coastal Ranges, provide a unique empirical baseline covering distinct lithologies, varying block geometries, and a wide spectrum of impact energies. By systematically calibrating the model against laboratory tests and anchoring it in these documented field events, we perform an extensive parametric exploration to evaluate how fragmentation intensity and debris evolution emerge from the interplay between gravitational potential energy and intrinsic rock strength. The results are expressed and interpreted through a dimensionless framework, utilising length ratios and relative breakage indices to normalise the fragmentation response across different scales. This approach seeks to identify generalised physical principles and scaling regularities that govern brittle failure, providing new quantitative constraints for linking discrete impact mechanics with macroscopic deposit characteristics in complex geological terrains.

\section*{The Catalonia rockfall database: a natural laboratory for fragmentation}

\begin{figure}[htbp]
\centering
\includegraphics[width=1\linewidth]{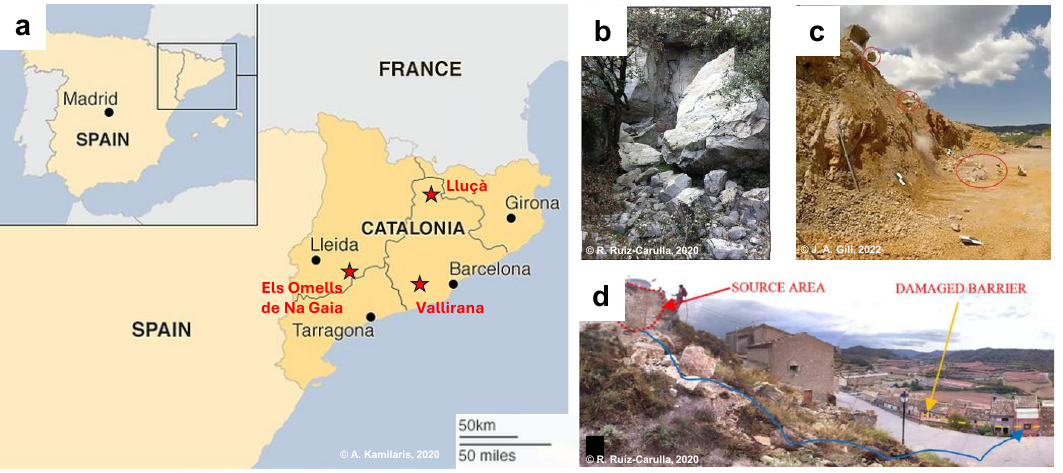}
\caption{Reported rockfall events on which this study is based \cite{Ruizcarulla2016,Gili2022}. (a) Location of events in Catalonia, Spain. (b) Rockfall in resistant cemented sandstone, Lluçà. (c) Limestone open pit in Vallirana. (d) Event in sedimentary units in Els Omells de Na Gaia.}
\label{fig:events}
\end{figure}

To anchor our numerical framework in real-world scenarios, we selected three high-resolution case studies from Catalonia, Spain, which represent a diverse lithological and geomechanical baseline (Fig. \ref{fig:events}a). These sites have been the subject of extensive in-situ experimental campaigns and detailed documentation, providing a robust dataset for model validation:

\begin{itemize}
    \item Lluçà (High-strength siliciclastic unit, Fig. \ref{fig:events}b): A high-energy event involving competent, well-cemented sandstone blocks. This case represents the high-strength end-member of the sedimentary spectrum, providing a rigorous test for the model’s ability to resolve fragmentation in highly resistant siliciclastic lithofacies.
    \item Vallirana (Carbonate sequence, Fig. \ref{fig:events}c): An experimental quarry characterised by massive Mesozoic limestones, where real-scale drop tests were conducted to track block trajectories and fragmentation patterns \cite{Gili2022}.
    \item Els Omells de na Gaia (Low-strength siliciclastic unit, Fig. \ref{fig:events}d): A rockfall event involving Paleogene sandstones from the Ebro Basin. This site offers insights into the disaggregation of poorly to moderately cemented sedimentary units with lower intact rock strength.
\end{itemize}

\begin{table}[hbp]
\centering
\caption{\label{tab:reported}Global characteristics of inventoried rockfalls.}
\begin{tabular}{llll}
\toprule
Parameter & Case I & Case II & Case III \\
\midrule
Location & Vallirana & Omells & Lluçà \\
Failure mechanism & Controlled & Slide & Toppling \\
Material & Limestone & Low-strength sandstone & Sandstone \\
Total volume RBSD (m$^3$) & 0.5 & 4.2 & 10.7 \\
Total volume IBSD (m$^3$) & 0.5 & 4.2 & 10.7 \\
RBSD total nº of estimated blocks & 57 & 48 & 78 \\
RBSD nº of measured blocks & 63 & 48 & 78 \\
Min. measured vol. (m$^3$) & 0.00002 & 0.0007 & 0.0007 \\
Max. measured vol. (m$^3$) & 0.2 & 1.1 & 8.47 \\
Total impact height (m) & 16.5 & 14.5 & 6.6 \\
Run-out distance (m) & 5 & 22 & 7.5 \\
\bottomrule
\end{tabular}
\end{table}

These events are particularly significant as they provided the empirical basis for the Fractal Fragmentation Model (FFM) \cite{Ruizcarulla2015,Ruizcarulla2017}. While the FFM successfully describes the power-law nature of rockfall deposits, our study aims to extend this understanding by explicitly resolving the physical mechanisms of breakage. By integrating these well-documented events with discrete numerical simulations, we bridge the gap between fractal descriptions and the energy-driven mechanics that govern the emergence of universal scaling laws. Table \ref{tab:reported} shows the characteristics and properties of these events, including the volumes involved and their run-out distance.

\section*{Description of the numerical framework}
\subsection*{Energy-based fracture criterion}

The physics of discrete and particulate systems is often modelled by directly solving the Newton-Euler equations of motion (translation and rotation of rigid bodies), which allow to accurately determine the position of each body at different instants in time. However, dynamic phenomena involving granular media fragmentation require an extension to kinematic models. In this study, the fracture is modelled through a breakage expansion (the fast-breakage method, FBM \cite{Ansys2025}), a type of particle replacement model that allows the instantaneous breakage of a particle, replacing it with progeny fragments formed by irregular polyhedral \cite{Paluszny2016}. This model applies the extended breakage theory, based on fracture mechanics models according to its statistical behaviour and similarity reasoning of fracture schemes \cite{Vogel2003,Vogel2004,Shi2007}.

In this widely used model, fragmentation occurs when the impact energy experienced by a particle exceeds a critical threshold, the material's minimum fracture energy, leading to its substitution by smaller sub-particles. Based on this principle, we hypothesise the occurrence of material fracture can be described as a probabilistic variable as a continuous function of impact energies:
\begin{equation}
    P_{b}=1- \exp \left[- S E_{cum} \left (\frac{d_i}{d_{ref}} \right) \right]\
    \label{eq:heavisideprob}
\end{equation}

\noindent in which:
\begin{equation}
    E_{cum}={E}'_{cum}+E-E_{min}
\label{Eq7:ecum}
\end{equation}
\[
    E_{min}=E_{min,ref}\left ( \frac{d_i}{d_{ref}} \right)
\]

\noindent where $S$ is a measure of the fracture resistance of the material (known as the selection function, in kg/J), ${E}'_{cum}$ is the accumulated energy prior to a stressing event, $E$ the specific impact energy, $d$ and $d_{ref}$ the sizes of the impacted and the reference particle, respectively, and $E_{min,ref}$ the minimum impact energy sufficient to cause fracture in a particle of size $d_{ref}$. Thus, when the energy applied to the particle is greater than the fracture energy of the particle, it will break, generating progeny fragments, whose size distribution is obtained from the fracture index $t_{10}$ of the material:

\begin{equation}
    t_{10}=A\left \{1-\exp \left[-S E_{cum} \left ( \frac{d_i}{d_{ref}} \right )\right]\right\}
\label{Eq9:t10}
\end{equation}

\noindent where $t_{10}$ is the fineness index that represents the percentage of the initial mass of the particle that will pass through a sieve of 1/10th of the original size $d_i$, and $A$ is a fit parameter that describes the maximum $t_{10}$ for a particle subject to breakage, obtained through calibration in drop-weight tests.

The value of the fineness index of the material allows to obtain the size distribution of the material, with the help of the incomplete beta function \cite{Barrios2011,Barrios2013}:
\begin{equation}
    t_n=\frac{100}{\int_{0}^{1}x^{\alpha_n-1}(1-x)^{\beta_n-1}dx}\int_{0}^{t_{10}}x^{\alpha_n-1}(1-x)^{\beta_n-1}dx
\label{Eq10:incompletebeta}
\end{equation}

\noindent where $\alpha_n$ and $\beta_n$ are fitting coefficients of the model obtained through calibration in impact tests. In this way, the model allows the complete description of the grain size distribution knowing the characteristics of the $t_n$-family curves. An application of this probability approach shows that the survival rate of a rock block is closely related to the geometry of the system \cite{Buzzi2023,Vergara2025}. Fig. \ref{fig:fbm}a shows the calculation scheme of the breakage model, as well as the generation of progeny particles after impact.

\subsection*{Spatial tessellation and creation of new fragments}

The internal topology of the rock blocks was generated using a Laguerre-Voronoi tessellation (or Dirichlet cell complex) \cite{Imai1985,Du2002}. The generator points are seeded following a stochastic distribution where weights are assigned to match the experimental particle size distribution, ensuring a space-filling polyhedral assembly without gaps. Formally, given a convex domain $\Omega\subset \mathbb{R}^3$, $n$ distinct generator points: $x_1,…,x_i\in \Omega$ and corresponding weights (inversely proportional to impact energy) $w_1,…,w_n\in\mathbb{R}$, the Laguerre-Voronoi diagram $\left \{ L_i \right \}_{i=1}^n$ generated by $(x_1,w_1 ),…,(x_n,w_n)$ $ \forall j\in {1,...,n}$ is defined by:
\begin{equation}
    L_i=\left \{ x\in\Omega :||x-x_i||^2-w_i\leq ||x-x_j||^2-w_j \right \}
\label{Eq11:voronoi}
\end{equation}

\begin{figure}[hbp]
\centering
\includegraphics[width=0.5\linewidth]{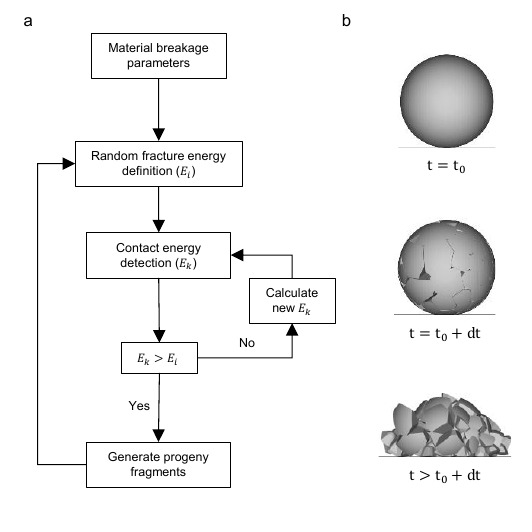}
\caption{Calculation scheme of the fast-breakage numerical model. \textbf{a} Work sequence of the particle fracture model. \textbf{b} Visual diagram of the replacement fragmentation process of a smoothed spherical particle.}
\label{fig:fbm}
\end{figure}

\noindent This algorithm is further characterised by the generation of the smallest fragments in the vicinity close to the contact point, and larger fragments far from this vicinity (power diagram), more realistically imitating the brittle fracture of rock materials (Fig. \ref{fig:fbm}b).

To implement the discrete breakage framework described above, throught the FBM framework, we employed the high-performance Ansys Rocky software package. This platform was selected due to its advanced capability to model non-spherical, polyhedral particles, which more accurately represent the angularity and interlocking nature of real-world rock blocks compared to traditional multi-sphere approaches.

\subsection*{Breakage Quantification}

\begin{figure}[hbp]
\centering
\includegraphics[width=0.45\linewidth]{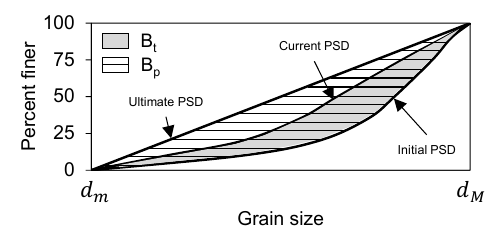}
\caption{Quantification of breakage based on the areas enclosed by the different size distribution curves \cite{Einav2007}.}
\label{fig:einav_def}
\end{figure}

Among the proposals for quantifying the degree of fragmentation, one of the most conventional is based on the comparison of the size distribution curves before and after a fracture event, from which a maximum cut-off value is defined above which a particle cannot undergo further fragmentation \cite{Hardin1985}. However, a disadvantage of this methodology is that it assumes that the materials will reach maximum breakage when the component grains of the granular medium are of a size equal to 75 $\mu$m, which is reached by certain materials under highly stressful environments. To provide a consistent definition, a modification to the estimation of the relative breakage, $B_r$, is proposed (Fig. \ref{fig:einav_def}), from which the proximity of the current size distribution curve between an initial and a final one is derived:
\begin{equation}
    B_r=\frac{\int_{d_m}^{d_M}(F_c(d)-F_0(d))d^{-1}\textrm{d}d}{\int_{d_m}^{d_M}(F_u(d)-F_0(d))d^{-1}\textrm{d}d}
    \label{eq:definition}
\end{equation}

\noindent where $F_0(d)$ is the initial size distribution (equivalent to the in-situ block size distribution), $F_c(d)$ is the current size distribution (equivalent to the rockfall block size distribution), and $F_u(d)$ the ultimate size distribution, defined as the maximum distribution curve that a material can reach under intense fracture conditions. We consider the latter to be of utmost importance, as it allows us to narrow down the definition of breakage. Thus, the area defined between the initial fragmentation curve (IBSD) and $F_u(d)$ is defined as the potential breakage, $B_p$, a key parameter in the search for a unified model.

\subsection*{Calibration of micro-mechanical properties via numerical drop-weight tests}

\begin{figure}[htbp]
    \centering
    \includegraphics[width=0.65\linewidth]{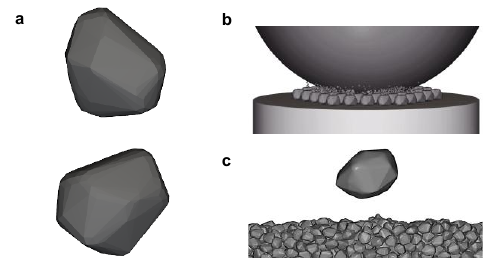}
    \caption{\textbf{a} Irregular rock geometry used in the study. \textbf{b} DWT scheme for numerical calibration tests. \textbf{c} Diagram of the rockfall used in the numerical model, based on the free fall of a rock block impacting onto a granular bed.}
    \label{fig:geometry}
\end{figure}

The micro-mechanical and breakage properties of the selected lithotypes were determined by replicating the experimental protocol of the Drop-Weight Test (DWT) as described in the benchmark studies of Jiménez Herrera et al. \cite{JimenezHerrera2018} and Tavares et al. \cite{Tavares2020}. In these numerical experiments, an 88-mm diameter steel projectile was released onto discrete particles of each lithology, positioned atop a fixed steel anvil to ensure precise energy transfer. Impact energies were systematically modulated by varying the drop height, achieving a controlled velocity range between 1 and 2.5 m/s. To ensure statistical robustness and capture the stochastic nature of brittle failure, each lithology underwent 15 to 20 independent trials per energy level. Rock blocks are represented by irregular geometries, based on the fractal characteristics of spherical harmonics \cite{Wei2018}. This morphology allows for a faithful and realistic representation of the rock's mechanical behaviour, avoiding the restrictions imposed by discrete models using spherical or regular geometries  (Fig. \ref{fig:geometry}a,b).

\begin{figure}[htbp]
\centering
\includegraphics[width=0.9\linewidth]{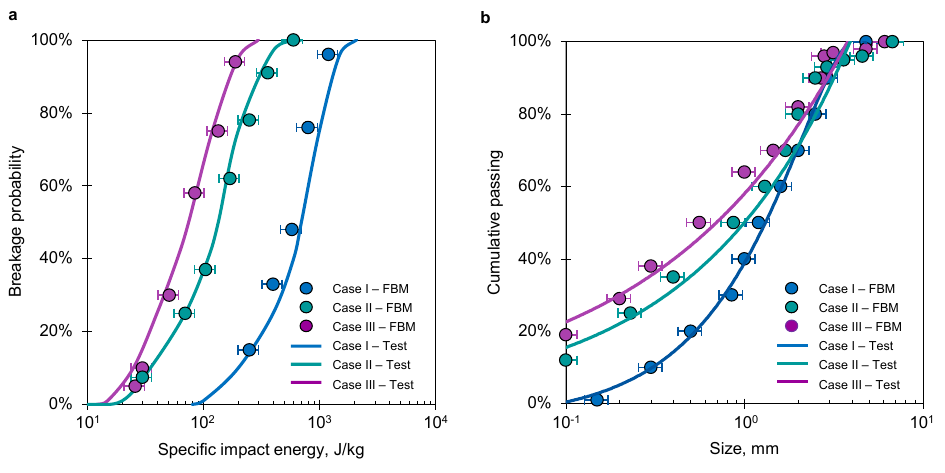}
\caption{Results of the calibration process and comparison between the experimental and numerical results \cite{JimenezHerrera2018,Tavares2020}. \textbf{a} Breakage probability as a function of the specific impact energy applied to the three types of materials. \textbf{b} Size distribution curve for an impact energy of 3 J.}
\label{fig:calibration}
\end{figure}

The calibration objective was to match two fundamental indicators: the fragment size distributions (FSD) and the breakage probability ($P_b$) as a function of specific impact energy. Here, $P_b$ is defined as the ratio of successful fracture events to the total number of trials at a given energy threshold. The resulting alignment between experimental benchmarks and our numerical simulations via FBM (Fig. \ref{fig:calibration}a,b) provided the calibrated input parameters for the breakage model (summarised in Tables \ref{tab:characteristics} and \ref{tab:breakage}). These parameters, specifically the $\alpha$ and $\beta$ coefficients governing the energy-size relationship \cite{Barrios2011}, were subsequently kept constant during the validation phase to test the model's predictive performance.

\begin{table}[btp]
\centering
\caption{\label{tab:characteristics} Mechanical characteristics of the tested material.}
\begin{tabular}{llll}
\toprule
Parameter & HS Sandstone & Limestone & LS Sandstone \\
\midrule
\textit{Material parameters}   & & & \\
Young's modulus (GPa)          & 52   & 20   & 19.9  \\
Poisson's ratio                & 0.25 & 0.25 & 0.33  \\
Density (kg/m$^3$)             & 2930 & 2710 & 2650  \\
\textit{Interactions}          & & & \\
Coefficient of static friction & 0.16 & 0.39 & 0.49 \\
Coefficient of restitution     & 0.38 & 0.48 & 0.33  \\
\bottomrule
\end{tabular}
\end{table}

\subsection*{Validation against field events and parametric scaling}

The predictive robustness of the discrete element framework was validated by simulating the three well-documented rockfall events in Catalonia. These scenarios provide a rigorous "blind test" for the model, as they encompass diverse lithologies and failure mechanisms. For each case, the In-situ Block Size Distribution (IBSD) was reconstructed using Discrete Fracture Network (DFN) analysis and field characterisation of the source scars \cite{Gili2022}. 

\begin{figure}[htbp]
\centering
\includegraphics[width=0.5\linewidth]{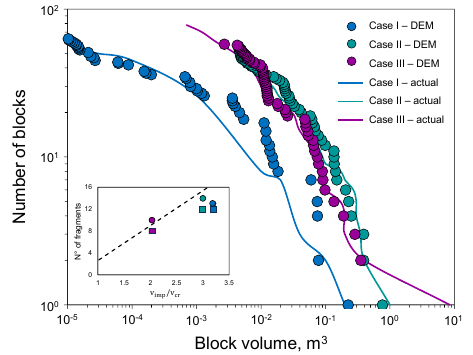}
\caption{Comparison between actual and modelled fragments based on the reported events in Catalonia \cite{Ruizcarulla2020,Gili2022}. The dots represent the volumes of the individual fragments after the rockfall, modelled for each case. Solid lines represents the volume of the blocks after impact, estimated from field visualisation. Note that the RBSD is obtained from the fracture of all individual impactor blocks (IBSD). The internal graph shows the number of blocks (>5\% of the initial size) as a function of the normalised impact velocity, compared between what was reported (square markers), what was modelled (circular markers), and the predictive model (dashed line) \cite{Guccione2025}.}
\label{fig:validation}
\end{figure}

Validation was performed by comparing the simulated Rockfall Block Size Distributions (RBSD) against the field observations reported in Table \ref{tab:reported}. As shown in Fig. \ref{fig:validation}, the model accurately captures the transition from the IBSD to the fragmented RBSD, reflecting the material-specific response to kinetic energy dissipation.

Based on the calibrated parameters shown in Tables \ref{tab:characteristics} and \ref{tab:breakage}, for each of the materials tested, a comprehensive parametric analysis was conducted to explore the sensitivity of fragmentation to initial conditions. Simulations consider the free fall of rock blocks onto smooth surfaces, represented by a granular bed of particles sized $D/10$, extending beyond $20D$ to minimise boundary effects (Fig. \ref{fig:geometry}c). We systematically varied the fall height ($h$) from 1 to 200 metres and the initial block diameter ($D$) from 0.5 to 4 metres across the three lithological scenarios (see Table \ref{tab:simulations}). This multi-scale approach allows for the observation of how breakage efficiency and fragment size spectra evolve as emergent properties of the interplay between gravitational potential energy and intrinsic rock strength. These parameter ranges were strategically selected to reflect the natural variability observed in Mediterranean and Alpine rockfall events. A maximum height of 200 m characterises high-energy trajectories in steep rock walls, while the block diameters ($0.5 \leq D \leq 4.0$ m) represent the most frequent sizes identified in the field characterisation of source scars. By exploring this extensive parametric space, we ensure that the emergent statistical regularities—specifically the transition from initial integrity to pervasive fragmentation—are captured across several orders of magnitude of impact energy.

\begin{table}[htbp]
\centering
\caption{\label{tab:breakage} Energy parameters of the numerical model for each type of material.}
\begin{tabular}{llll}
\toprule
Parameter & LS sandstone & Limestone & HS Sandstone \\
\midrule
Reference size (m)                    & 0.005 & 0.011 & 0.003  \\
Minimum specific energy (J/kg)        & 213.5 & 15.8  & 43.6  \\
Selection function coefficient (kg/J) & 0.020 & 0.033 & 0.018  \\
Maximum $t_{10}$ value     & 0.677 & 0.533 & 0.388  \\
$\alpha_{1.2}/\beta_{1.2}$ & 0.505/11.95 & 0.155/6.219 & 0.100/15.39 \\
$\alpha_{1.5}/\beta_{1.5}$ & 1.066/13.87 & 0.397/5.468 & 0.346/8.090 \\
$\alpha_{2}/\beta_{2}$     & 1.014/8.088 & 0.770/5.538 & 0.872/8.437 \\
$\alpha_{4}/\beta_{4}$     & 1.084/3.027 & 1.106/3.076 & 0.885/2.213 \\
$\alpha_{25}/\beta_{25}$   & 1.012/0.527 & 1.165/0.540 & 1.048/0.430 \\
$\alpha_{50}/\beta_{50}$   & 1.026/0.363 & 1.481/0.413 & 1.075/0.210 \\
$\alpha_{75}/\beta_{75}$   & 1.034/0.295 & 1.776/0.365 & 1.060/0.137 \\
\bottomrule
\end{tabular}
\end{table}

\begin{table}[htbp]
\centering
\caption{\label{tab:simulations}Input geometric parameters for the simulation set.}
\begin{tabular}{lll}
\toprule
Parameter            & Range          & Unit \\
\midrule
Rock block size,$D$  & [0.5, 1, 2, 4] & [m] \\
Rockfall height, $h$ & [1, 2, 4, 10, 25, 50, 100, 200] & [m]  \\
\bottomrule
\end{tabular}
\end{table}

\newpage
\section*{Results and discussion}
\subsection*{Scaling and emergence of a fragmentation signature}

To ensure dynamic similarity across varying scales and lithologies, we adopt the dimensionless length ratio, $\ell \equiv h/D_i$, as the primary governing parameter. Unlike traditional empirical proxies such as run-out distance, which are heavily influenced by extrinsic topographical constraints, the length ratio provides a scale-invariant measure of impact intensity. This parameter, whose robustness for quantifying rockfall fragmentation was previously established \cite{Vergara2025}, effectively normalises the gravitational potential energy relative to the characteristic dimension of the block. By doing so, $\ell$ serves as a proxy for the energy density available for comminution at the moment of impact. This approach facilitates the identification of scaling laws, as it isolates the intrinsic mechanical response of the rock mass from the absolute magnitudes of fall height and block volume, thereby ensuring that the fragmentation trends observed are physically scalable.

\begin{figure}[hbp]
\centering
\includegraphics[width=0.5\linewidth]{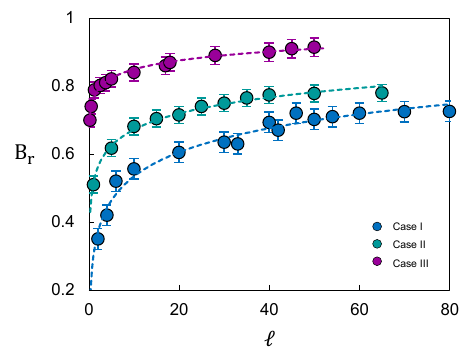}
\caption{Relative breakage, $B_r$, versus length ratio, $\ell$. The dashed lines represent the exponential fit to each set of points.}
\label{fig:quantif}
\end{figure}

Fig. \ref{fig:quantif} illustrates the evolution of the relative breakage, $B_r$, as a function of the length ratio $\ell$ for the three studied lithotypes. Despite the disparate geomechanical properties of the materials, the fragmentation response exhibits a consistent phenomenological trend. The data follow an exponential saturation profiles described by $y=a(1-e^{-bx})$, where the coefficients $a$ and $b$ act as lithological-specific sensitivities. This behaviour reveals two distinct regimes: (i) an initial stage of rapid, energy-sensitive breakage, followed by (ii) a fragmented plateau where additional kinetic energy yields diminishing returns in comminution—a clear manifestation of the energetic limits of brittle fragmentation.

\begin{figure}[bp]
\centering
\includegraphics[width=0.5\linewidth]{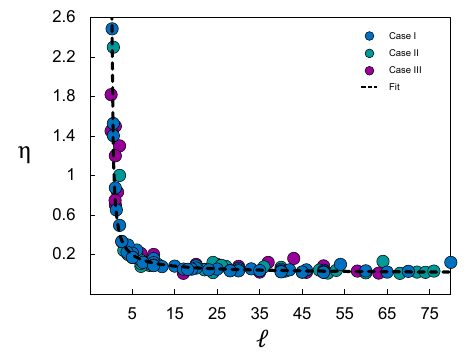}
\caption{Collapse of the data set into a dimensionless system. The breakage ratio is herein defined as the ratio of relative breakage to potential breakage. The solid curve corresponds to the best fit to the data, representing the fragmentation prediction model (Eq. \ref{eq:proposal}).}
\label{fig:dimensionless}
\end{figure}

To identify a global behaviour, we developed a Normalised Breakage Index, $\eta \equiv B_r/[1-\exp(-B_p)]$. This index acts as a scaling kernel that accounts for the potential breakage $B_p$, thereby isolating the intrinsic efficiency of the fragmentation process. When the simulated data are projected onto this dimensionless space (Fig. \ref{fig:dimensionless}), a remarkable phenomenon occurs: the scattered data points from all three lithotypes collapse onto a single master curve. This data collapse suggests that rockfall fragmentation is governed by a robust, scale-invariant law, expressed as:
\begin{equation}
    \eta = \kappa_1 \ell^{\kappa_2}
    \label{eq:proposal}
\end{equation}

In this expression, the constants $\kappa_1 = 0.32$ and $\kappa_2 = -0.15$ represent the fundamental scaling exponents obtained through non-linear regression ($R^2>0.9877$). The mathematical structure of Eq. \ref{eq:proposal} ensures that the predicted relative breakage remains bounded within the physical limits $[0, 1]$, preventing non-physical over-prediction at extreme energy regimes. 

\begin{figure}[hbp]
\centering
\includegraphics[width=0.5\linewidth]{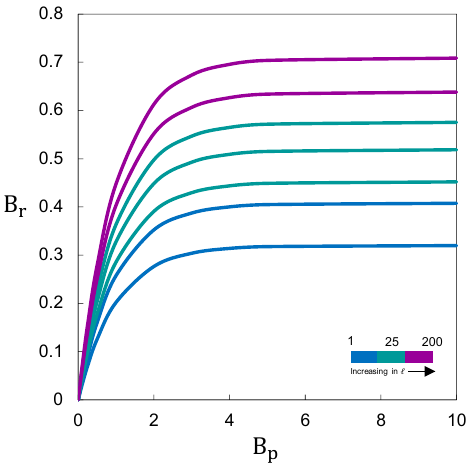}
\caption{Sensitivity analysis of relative breakage as a function of potential breakage. The family of curves is created based on the proposed equation, for different values of $\ell$.}
\label{fig:sensitivity}
\end{figure}

The convergence of diverse lithological responses into this unified signature (Fig. \ref{fig:sensitivity}) suggests that, although material strength dictates the magnitude of breakage, the statistical regularities of the fragmentation process are intrinsically linked to the impact dynamics. This framework effectively bridges the gap between discrete mechanical failure and macroscopic debris evolution, providing a predictive tool that integrates the complex interplay of impact height, block geometry, and lithological inheritance.

While fractal-based descriptions provide a static representation of the final deposit, our dimensionless approach explicitly resolves the dynamic evolution of the breakage process. The identified scaling law complements fractal theory by providing the energetic context required to predict how a specific IBSD transitions into a fragmented RBSD under varying impact conditions.

\newpage
\subsection*{Statistical analysis based on Weibullian properties}

Following Weibull's theory of material strength \cite{Weibull1951}, fragmentation events are governed by this type of statistical distribution of impact energies. In the following, we will extend this hypothesis. 
Let $m$ and $\lambda$ be the shape and scale parameters, respectively; the Weibull probability density function $f$ for relative breakage is then defined as:
\begin{equation}
    f(B_r)=\frac{m}{\lambda}\left( \frac{B_r}{\lambda} \right)^{m-1}\exp\left[ -\left( \frac{B_r}{\lambda} \right)^m \right]
\end{equation}

By integrating the probability density function between 0 and $B_r$, the cumulative distribution function is obtained as follows:
\begin{equation}
    F(B_r)=\int_{0}^{B_r} f(t)dt=1-\exp\left[ -\left( \frac{B_r}{\lambda} \right)^m \right]
\end{equation}

The scale parameter $\lambda$ can be understood as the characteristic breakage $B_{r,0}$, the value at which the relative breakage reaches $\sim$ 37\% probability. Thus, we define the survival probability $P_s$ as the complement of the cumulative distribution function:
\begin{equation}
    P_S=1-F(B_r)=\exp\left[ -\left( \frac{B_r}{\lambda} \right)^m \right]
\end{equation}

Finally, linearising we get:
\begin{equation}
    \ln \left[ \ln(1/P_S) \right]=m\ln B_r-m \ln B_{r,0}
\end{equation}

\begin{figure}[htbp]
\centering
\includegraphics[width=1\linewidth]{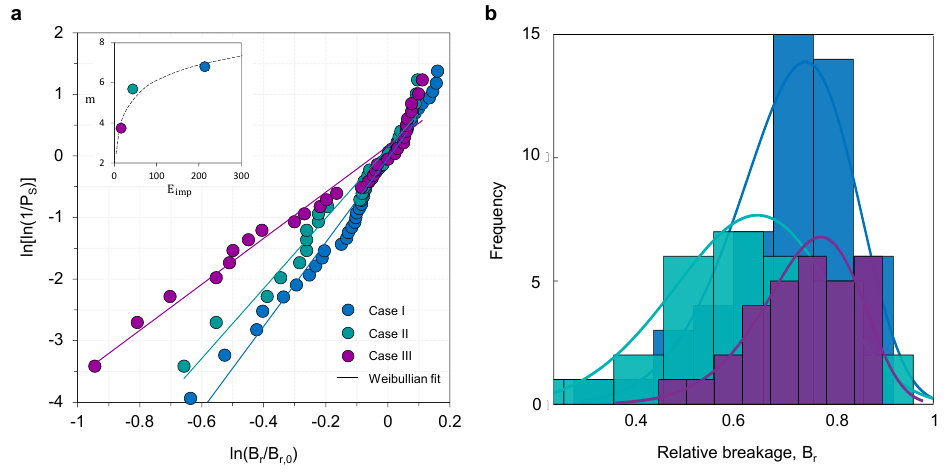}
\caption{Weibull analysis of rockfall fragmentation. \textbf{a} Weibull breakage distribution for the different data sets. The inner graph shows the behaviour of the parameter $m$ as a function of the specific impact energy for each material. \textbf{b} Set of frequency histograms for each data set. The solid lines represent the fit using the Weibull distribution.}
\label{fig:weib}
\end{figure}

Based on the re-plotting of the results shown in Fig. \ref{fig:quantif}, Fig. \ref{fig:weib}a presents a linearisation of the Weibull distribution, in which we define the characteristic breakage, $B_{r,0}$, as the breakage value under the assumption of $P_S=37\%$ (also known as the scale parameter of the distribution). The linear fits for the three cases show a high correlation ($R^2 > 0.9664$), allowing us to support our hypothesis. The internal graph in Fig. \ref{fig:weib}a represents the values of the curve slopes, the distribution parameter $m$. Considering that the Weibull modulus $m$ reflects the statistical variability of the data, it is reasonable to infer that materials subjected to higher specific impact energies tend to display lower dispersion in fragment sizes. This suggests that fragmentation becomes more predictable in mechanically stronger rocks, whereas in lower-strength lithologies, the process is inherently more stochastic—potentially making the model less sensitive or more prone to overestimating breakage under highly brittle conditions.

The histograms in Fig. \ref{fig:weib}b show a high-quality of fit and exhibit the characteristic asymmetry and decay behaviour expected for Weibull distributions, supporting the hypothesis that fragmentation is governed by the statistical distribution of impact energies.

The observed fragmentation patterns consistently adhere to a Weibull form, reinforcing the notion that brittle failure arises from the probabilistic activation of flaws under stress. This statistical behaviour is widely documented in ceramics and glasses, where scaling emerges from the heterogeneous distribution of microstructural defects \cite{Turcotte1986,Yashima1987,Hu2025}.
It is worth noting that while our discrete element framework does not explicitly resolve internal crack propagation—relying instead on an energy-based stochastic breakage model—the strong agreement with field and experimental data suggests that the framework effectively captures the macroscopic manifestation of these underlying physical processes. By utilising a breakage probability function that acts as a surrogate for the flaw-density distribution within the rock mass, the model reproduces these signatures without the computational overhead of explicit fracture mechanics.

The consistency of this pattern across material classes and scales suggests that Weibull fragmentation may reflect a universal law governing the breakage of brittle matter under dynamic loading. This universality provides a conceptual bridge between geological processes and material physics, and offers a predictive framework for understanding fragmentation in both natural and experimental systems. Consequently, the ability of the stochastic approach to converge on the same statistical signatures observed in more complex, crack-resolved systems highlights its robustness for large-scale geomorphic applications. Nevertheless, future work integrating discrete fracture networks (DFN) could further refine the link between lithological inheritance and the kinematics of fragment generation, providing an even more granular view of the failure process.

\newpage
\section*{Conclusions}

This study establishes a high-resolution numerical framework to decipher the mechanical and statistical principles governing rockfall fragmentation. By integrating Discrete Element Method simulations with high-fidelity field data from three distinct geomechanical settings in Catalonia, we provide a robust and transferable approach to quantifying brittle breakage in gravity-driven systems. Our findings lead to the following key conclusions:
\begin{itemize}
    \item Fragmentation laws: despite the disparate geomechanical properties of the studied lithotypes, the fragmentation response follows a consistent and predictable scaling behaviour. When normalised through dimensionless variables, the breakage data collapse onto a single signature. This reveals that while lithological inheritance dictates the magnitude of the response, the underlying physics of energy dissipation is scale-invariant.
    \item The Weibullian signature: as expected, our statistical analysis confirms that fragment size distributions consistently conform to a Weibull scaling law. This reinforces the hypothesis that brittle fragmentation in rockfalls is not a chaotic outcome, but a fundamental statistical regularity. Identifying this signature provides a new theoretical baseline for interpreting impact-induced failure across diverse geological scales.
    \item Predictive power for hazard assessment: the proposed framework explicitly resolves the partitioning of impact energy between the residual integrity of surviving blocks and the generation of comminuted debris. This capacity to link intrinsic rock strength with final deposit characteristics offers a transformative tool for improving the reliability of run-out modelling and hazard zoning in complex mountainous terrains.
    \item Geomorphic and landscape implications: Beyond its immediate engineering utility, the identification of a universal fragmentation law provides critical insights for the Earth sciences. If Weibullian fragmentation is an intrinsic trait of brittle geomaterials, it implies that the sedimentological characteristics of steep terrains are governed by robust statistical principles. The efficiency of this fragmentation process directly influences sediment connectivity, the morphometry of talus slopes, and the size distribution of debris delivered to hillslope–channel interfaces. By quantifying the transition from intact blocks to fragmented spectra, our work highlights rockfall fragmentation as a fundamental control on the initial state of sediment supply, bridging the gap between discrete mechanical failure and the broader geomorphic processes that shape rocky landscapes.
    \item Design implications in civil engineering: the ability to predict fragment size distributions through a physics-based scaling law represents a paradigm shift for rockfall mitigation. By moving beyond the \textit{single-block} assumption, engineers can now evaluate the energy partitioning of fragmented clouds, optimising the design of protection galleries and the mesh apertures of flexible barriers. This framework provides a standardised metric—the material breakage susceptibility—to classify rock slopes not only by their breakage probability but by their fragmentation potential, a critical factor for the structural resilience of infrastructure in mountainous regions.
\end{itemize}

While this study offers a unified perspective on rockfall dynamics, it also highlights future research avenues. The current computational constraints on extreme block sizes and the inherent stochasticity of natural terrain roughness suggest that integrating more complex boundary conditions will be the next frontier. Expanding this framework to include multi-impact sequences, pre-existing flaw density and heterogeneous rock masses will further consolidate its applicability as a generalised principle for characterising the dissipative nature of our planet's most dynamic landscapes.

\section*{Data availability}

The list of simulation results, calibration and validation steps, as well as any analysis and post-processing databases, are available from the corresponding author upon reasonable request.

\newpage
\bibliography{library}

\section*{Acknowledgements}

The authors thanks the sponsorship of the \textit{Deutscher Akademischer Austauschdienst} (DAAD), and RWTH Aachen for providing the software licensing and computational infrastructure required to perform the numerical simulations presented in this study. Á.V. thanks the Dr. Wei for providing support in the generation of geometries applied in this study.

\section*{Author contributions}

Á.V. implemented and optimised the numerical breakage model, obtained data from the literature, performed data curation, and mathematically processed the information obtained in each simulation. The review of the results and the design of figures was supported by S.P. The project administration, main conception of the study, revision, interpretation of results, and main supervision were guided by R.F. Á.V. wrote the main manuscript, based on contributions from all authors.

\section*{Funding}

This work is supported by the National Scientific Research Agency of Chile (ANID), Grants Nos 57600326 and 1211469.

\section*{Competing interests}

The authors declare no competing interests.



\section*{Additional information}


\noindent \textbf{Supplementary information} and requests for materials should be addressed to Álvaro Vergara.

\noindent \textbf{Correspondence} and requests for materials should be addressed to Álvaro Vergara.

\end{document}